\def\be{\begin{equation}}
\def\ee{\end{equation}}
\def\bea{\begin{eqnarray}}
\def\eea{\end{eqnarray}}

\documentclass[aps,prb,twocolumn,amsmath,amssymb]{revtex4}
\usepackage{epsfig,graphicx}
\usepackage{bm}

\begin{document}


\title{Quantum efficiency of binary-outcome detectors of solid-state qubits}

\author{Alexander N. Korotkov}
\affiliation{Department of Electrical Engineering, University of
California, Riverside, CA 92521-0204, USA}

\date{\today}


\begin{abstract}
We discuss definitions of the quantum efficiency for binary-outcome
qubit detectors with imperfect fidelity, focusing on the subclass of
quantum non-demolition detectors. Quantum efficiency is analyzed for
several models of detectors, including indirect projective
measurement, linear detector in binary-outcome regime, detector of
the superconducting phase qubit, and detector based on tunneling
into continuum.
\end{abstract}

\maketitle

\section{Introduction}

   Reliable measurement of qubits is an essential requirement for
the operation of a quantum computer. \cite{Bennett,Nielsen} A
perfect detector of the qubit state should perform projective
measurement, \cite{Neumann} which means that it should have two
possible outcomes: 0 and 1 (corresponding to the qubit states
$|0\rangle$ and $|1\rangle$), the probabilities of these outcomes
should be equal to the corresponding matrix elements of the qubit's
density matrix (traced over other entangled qubits), and after the
measurement the pre-measured quantum state should get projected onto
the subspace, corresponding to the measurement result. Realistic
detectors of course do not realize the perfect projective
measurement. In this paper we discuss efficiency of a realistic
detector, assuming for simplicity the measurement of only one qubit;
in other words, we neglect physical coupling with other entangled
qubits in the process of measurement, so that the many-qubit
measurement problem can be reduced to the one-qubit measurement.

    We will consider a detector with binary outcome: 0 or 1 (such
detector can also be called a dichotomic or threshold detector). In
a realistic case the outcome does not perfectly correspond to the
qubit state. For example, for a qubit in the state $|0\rangle$ the
probability $F_0$ of the result 0 is typically less than 100\% (we
will call $F_0$ a ``measurement fidelity'' for the state
$|0\rangle$). Similarly, for a qubit in the state $|1\rangle$ the
probability $F_1$ to get result 1 is usually also less than unity.
(The fidelities $F_0$ and $F_1$ fully determine the outcome
probabilities for a general qubit state because of the linearity of
the quantum mechanics.) In such situation the measurement is surely
non-projective, and the post-measurement state should be analyzed
using a more general formalism. In this paper we discuss the quantum
efficiency (ideality) of binary-outcome detectors, which is defined
via decoherence of the post-measurement state. We emphasize that the
quantum efficiency is {\it not} related to the measurement
fidelities $F_0$ and $F_1$, and therefore is not directly related to
the fidelity of a quantum computer read-out. However, quantum
efficiency is an important characteristic of a detector with
imperfect fidelity, because for example, quantum-efficient detectors
can be used for implementation of non-unitary quantum
gates,\cite{Katz} quantum
feedback,\cite{Wiseman-Milburn,Geremia,Ruskov-02} quantum
uncollapsing,\cite{QUD,Katz-QUC} etc.

    Quantum efficiency (ideality) of solid-state qubit detectors has
been well studied for linear detectors.
\cite{Kor-99,Kor-01,Averin,Devoret-Sch,Makhlin-RMP,Pilgram,Clerk-03,Kor-nonid,Av-Sukh,Jordan-05,Clerk-06}
In this case it has been defined \cite{Kor-01} as $\eta = 1/2\Gamma
\tau_m$, where $\Gamma$ is the qubit ensemble decoherence rate due
to measurement and $\tau_m$ is the so-called ``measurement time'':
\cite{Shnirman-98} the time after which the signal-to-noise ratio
becomes equal to unity. Notice that the inequality $\Gamma \geq
1/2\tau_m$ (leading to the bound $\eta \leq 1$) can be easily
derived \cite{Kor-99} from the classical Bayes formula and
inequality $|\rho_{01}|^2\leq \rho_{00}\rho_{11}$ for the matrix
elements of the qubit density matrix $\rho$ (this derivation is
within the framework of the quantum Bayesian theory
\cite{Kor-99,Kor-01,Kor-rev}  describing individual realizations of
measurement); it has been also derived using the framework of the
ensemble-averaged theory of linear quantum detectors.
\cite{Averin,Devoret-Sch,Makhlin-RMP,Pilgram,Clerk-03} It is
important to mention that an equivalent definition of the
quantum-limited linear detector has been discussed more than two
decades ago \cite{Clarke-81,Caves-82,Danilov,Braginsky-book} in
terms of the ratio between the effective energy sensitivity and
$\hbar /2$. In a simple model, \cite{Kor-nonid} the detector
nonideality ($\eta <1$) can be caused by an additional coupling of
the qubit with dephasing environment which increases the back-action
noise, or by additional output noise (i.e.\ amplifier noise), or by
both contributions. Correspondingly, the detector efficiency $\eta$
can be interpreted as a ratio $\eta =\Gamma_{\min}/\Gamma$, where
$\Gamma_{\min}=1/2\tau_m$ is the ``informational'' limit on
decoherence, determined by a given rate of information acquisition,
or as a ratio $\eta =\tau_{m,\min}/\tau_m$, where
$1/\tau_{m,\min}=2\Gamma$ is the maximum possible rate of
information acquisition for a given value of back-action strength
$\Gamma$. Particular interpretation as well as the real physical
reason of the nonideality are irrelevant from the point of view of
qubit measurement.

    An ideal linear quantum detector ($\eta =1$) causes no
decoherence of the measured qubit in each realization of the
measurement (i.e.\ for each measurement outcome) in the sense that
initially pure qubit state remains pure in the process of
measurement; an example of such detector is the quantum point
contact (QPC). \cite{Kor-99} Moreover, a detector with $\eta =1$
should not introduce any change of the phase between amplitudes of
the states $|0\rangle$ and $|1\rangle$, except due to possible
constant shift of energy difference between the states $|0\rangle$
and $|1\rangle$. The proof of the last statement is rather
simple:\cite{Kor-99} if the phase change would depend on the
detector outcome, then averaging over the outcomes would lead to
strict inequality $|\langle \rho_{01}\rangle | < \langle
|\rho_{01}|\rangle$ and therefore to $\Gamma > 1/2\tau_m$. However,
there is a class of linear detectors, for which the qubit state does
not decohere for any measurement outcome, but nevertheless there is
an outcome-dependent phase shift between the states $|0\rangle$ and
$|1\rangle$.\cite{Stodolsky,Kor-Av,Goan-01,Averin,Pilgram,Clerk-03,Kor-nonid,Av-Sukh}
An example is an asymmetric QPC: in this case each electron passing
through the QPC shifts the phase between the qubit states by a small
constant.\cite{Av-Sukh} For such
detectors\cite{Kor-01,Kor-nonid,Averin,Devoret-Sch,Makhlin-RMP,Pilgram,Clerk-03}
$\Gamma = 1/2\tau_m +K^2S/4$, where $S$ is the output noise and $K$
is the properly normalized factor \cite{Kor-01,Kor-nonid} describing
the correlation between the output and back-action noises. Such
detectors are non-ideal by the above definition ($\eta <1$);
however, they are ideal in another sense: for example, they still
can be used for perfect quantum feedback, quantum uncollapsing, etc.
Therefore, it is meaningful to introduce a different definition of
the quantum efficiency which takes into account noise correlation
$K$, for example, as\cite{Kor-01,Kor-nonid} $\tilde \eta =
(1/2\tau_m )/(\Gamma -K^2S/4)$ or as $\tilde{\tilde \eta}=
(1/2\tau_m +K^2S/4)/\Gamma$ (here we exchanged the definitions of
$\tilde\eta$ and $\tilde{\tilde\eta}$ compared to Refs.\
\onlinecite{Kor-01,Kor-nonid,Kor-rev}).

    In the present paper we discuss various definitions of quantum
efficiency for binary-outcome detectors, using the reviewed above
methodology developed for the linear detectors. We start with
discussion of an arbitrary binary-oucome detector and show that in
general its quantum efficiency can be described by 18 parameters,
that is surely impractical. Then we focus on the class of detectors,
which do not affect qubit in the states $|0\rangle$ and $|1\rangle$
(we call them quantum non-demolition\cite{Braginsky-book} (QND)
detectors). Quantum efficiency of the QND detectors can be described
by only 2 parameters. We introduce several definitions of quantum
efficiency for the QND detectors and calculate the quantum
efficiencies for several detector models.

\section{General binary-outcome detector}\label{II}

    Let us start with the general description of the binary-outcome
measurement of a qubit, using the POVM (``positive operator-valued
measure'') theory of measurement.\cite{Nielsen} For an ideal
detector (which transforms a pure qubit state into a pure state) the
measurement can be described by two linear operators $M^{(0)}$ and
$M^{(1)}$, corresponding to two measurement results 0 and 1. For the
qubit with initial density matrix $\rho$ the probability of the
result 0 is $P_{0}=\mbox{Tr}(M^{(0)}\rho M^{(0)\dagger})$ and the
normalized post-measurement state for this result is $M^{(0)}\rho
M^{(0)\dagger}/P_{0}$ [quite often the non-normalized
post-measurement state $M^{(0)}\rho M^{(0)\dagger}$ is considered in
order to preserve the linearity of transformation]. Similarly, the
probability of the result 1 is $P_{1}=\mbox{Tr}(M^{(1)}\rho
M^{(1)\dagger})$ and then the post-measurement state is $M^{(1)}\rho
M^{(1)\dagger}/P_{1}$. The measurement operators $M^{(0)}$ and
$M^{(1)}$ should obey the following conditions:\cite{Nielsen} the
Hermitian operators $M^{(0)\dagger}M^{(0)}$ and
$M^{(1)\dagger}M^{(1)}$ should be positive (i.e.\ having only
non-negative eigenvalues) and satisfy the completeness relation
$M^{(0)\dagger}M^{(0)}+M^{(1)\dagger}M^{(1)}=1$. (The projective
measurement is a special case in which $M^{(0)}$ and $M^{(1)}$ are
projectors onto mutually orthogonal axes.)

    Let us count the number of degrees of freedom (real parameters)
describing such an ideal binary-outcome detector. A linear operator
$M^{(0)}$ acting in complex two-dimensional space can be described
by 4 complex numbers, i.e.\ 8 real parameters; however, one
parameter is the overall phase, so that there are 7 physical
parameters. Similarly, operator $M^{(1)}$ can be described by 7
parameters. The completeless relation gives 4 equations (two for
diagonal elements and two for the complex off-diagonal element).
Therefore, the measurement by an ideal binary-outcome detector can
be described by 10 real parameters (which include, in particular,
fidelities $F_0$ and $F_1$).

    A general (non-ideal) binary-outcome detector can be thought of as an ideal
detector with many possible outcome values, which however are
unknown to us, so that we know only if the outcome value belongs to
the group 0 or group 1 (this trick can obviously describe
information loss in an environment). Then the measurement can be
described by two groups of measurement operators $M^{(0)}_k$ and
$M^{(1)}_k$ with an extra index $k$ numbering operators within each
group. The probability of the result 0 in this case is $P_{0}=\sum_k
\mbox{Tr}(M^{(0)}_k\rho M^{(0)\dagger}_k)$ and the corresponding
post-measurement density matrix is $\sum_k M^{(0)}_k\rho
M^{(0)\dagger}_k/P_{0}$. Similar formulas can be written in the case
of result 1. The completeness relation in this case is $\sum_k
M^{(0)\dagger}_k M^{(0)}_k+ \sum_k M^{(1)\dagger}_kM^{(1)}_k=1$.
 Since separation
into measurement operators within each group is not unique, it is
better to deal with linear superoperators ${\cal S}^{(0)}$ and
${\cal S}^{(1)}$ (a superoperator transforms an operator into an
operator, i.e.\ a matrix into a matrix) defined as ${\cal
S}^{(0)}[\rho ]=\sum_k M^{(0)}_k\rho M^{(0)\dagger}_k$ and ${\cal
S}^{(1)}[\rho ]=\sum_k M^{(1)}_k\rho M^{(1)\dagger}_k$. In this
language the probability of the result 0 is $P_{0}= \mbox{Tr} \,
{\cal S}^{(0)}[\rho ]$ and the post-measurement state is ${\cal
S}^{(0)}[\rho ]/P_{0}$; the formulas are similar for the result 1.

    Now let us count the number of parameters describing a general
(non-ideal) binary-outcome measurement of a qubit. Following the
methodology of the quantum process tomography,\cite{Nielsen} we can
characterize the superoperator ${\cal S}^{(0)}$ by the result of its
operation on the state $|0\rangle$ (4 parameters, since the
resulting density matrix is non-normalized), the state $|1\rangle$
(4 more parameters), and operation on the non-physical density
matrix, for which one off-diagonal element is unity, while all other
elements are zero (this gives 8 more parameters, since resulting
matrix is not Hermitian). Overall this gives 16 parameters for
${\cal S}^{(0)}$ and similarly 16 parameters for ${\cal S}^{(1)}$.
The completeness relation gives 4 equations, therefore the total
number of remaining parameters is 28.

Comparing this number with 10 parameters for an ideal detector, we
see that quantum efficiency (or non-ideality) of a general
binary-outcome qubit detector should be described by 18 parameters.
This is surely impractical, and below we limit our discussion by a
narrower subclass of detectors, which can be described by a more
reasonable number of parameters.

\section{QND binary-outcome detector}\label{III}

    A general description of a detector attempted above does not
make any assumptions regarding the mechanism of measurement and is
based only on the linearity of the quantum mechanics. In particular,
this approach allows evolution of the qubit by its own in the
process of measurement (e.g.\ Hamiltonian evolution or energy
relaxation). Let us now make a restrictive assumption that the qubit
cannot not evolve by itself in the process of measurement. In
particular, we assume absence of coupling (infinite barrier) between
the measured states $|0\rangle$ and $|1\rangle$, so that coupling
with the detector can only affect the energy difference between
states $|0\rangle$ and $|1\rangle$ (in other words, we assume only
$\sigma_z$-type coupling). In this case the qubit initially in the
state $|0\rangle$ necessarily remains in the state $|0\rangle$ after
the measurement. Similarly, the qubit in the state $|1\rangle$
cannot evolve also. Such a detector is often called a QND detector
(the author does not quite like this terminology because it somewhat
differs from the original meaning of the quantum
non-demolition,\cite{Braginsky-book} but it will still be used here
because of absence of a better well-accepted terminology).

    First, let us consider an ideal QND detector, which transforms a
pure qubit state into a pure state. In the framework of the
methodology discussed in Sec.\ \ref{II}, we can characterize such
detector in the following way. In the case of result 0 the initial
qubit state $\alpha |0\rangle +\beta |1\rangle$ (here
$|\alpha|^2+|\beta|^2=1$) is transformed into the non-normalized
state $\alpha c^{(0)}_{0} |0\rangle +\beta c^{(0)}_{1}|1\rangle$
(normalization is trivial), while in the case of the result 1 it is
transformed into $\alpha c^{(1)}_{0} |0\rangle +\beta
c^{(1)}_{1}|1\rangle$. The probabilities of these results are,
correspondingly, $P_{0}=|\alpha c^{(0)}_{0} |^2 +|\beta
c^{(0)}_{1}|^2$ and $P_{1}=|\alpha c^{(1)}_{0} |^2 +|\beta
c^{(1)}_{1}|^2$, so that the measurement fidelities are
$F_0=|c^{(0)}_{0} |^2$ and $F_1=|c^{(1)}_{1} |^2$. The completeness
relation requires total probability of unity for any initial state:
$|c^{(0)}_{0} |^2+|c^{(1)}_{0} |^2=1$ and $|c^{(0)}_{1}
|^2+|c^{(1)}_{1} |^2=1$. Since the overall phases are not important,
we can assume, for example, that $c^{(0)}_{1}$ and $c^{(1)}_{1}$ are
real numbers. Overall, we have 4 parameters to characterize an ideal
QND detector: two fidelities ($F_0$ and $F_1$) and two phases
($\phi_0$ and $\phi_1$), so that in the case of result 0 the
wavefunction is transformed as
\begin{equation}
\alpha |0\rangle +\beta |1\rangle \rightarrow \frac{ \sqrt{F_0}\,
e^{i\phi_0}\alpha |0\rangle + \sqrt{1-F_1}\,\beta
|1\rangle}{\sqrt{P_{0}}} ,
    \label{transf0}\end{equation}
while in the case of result 1 the transformation is
\begin{equation}
\alpha |0\rangle +\beta |1\rangle \rightarrow \frac{ \sqrt{1-F_0}\,
e^{i\phi_1}\alpha |0\rangle + \sqrt{F_1}\, \beta
|1\rangle}{\sqrt{P_{1}}} ,
    \label{transf1}\end{equation} where
\begin{equation}
P_{0}=F_0 |\alpha|^2  +(1-F_1) |\beta|^2 , \,\, P_{1}=(1-F_0)
|\alpha|^2  +F_1 |\beta|^2 .
    \label{P_0,1} \end{equation}

    A non-ideal QND detector in general transforms a pure state into
a mixed state. Because of the linearity of superoperators $S^{(0)}$
and $S^{(1)}$ in the formalism discussed above, the only possible
modification of Eqs.\ (\ref{transf0}) and (\ref{transf1}) is an
extra decoherence between the states $|0\rangle$ and $|1\rangle$.
Therefore, a non-ideal binary-outcome QND detector can be
characterized by 6 parameters ($F_0$, $F_1$, $\phi_0$, $\phi_1$,
$D_0$, $D_1$), so that in the case of the result 0 the state
transformation is
\begin{equation}
\alpha |0\rangle +\beta |1\rangle \rightarrow \frac{1}{P_0} \left(
\begin{array}{cc} F_0 |\alpha |^2& \sqrt{F_0(1-F_1)}\, e^{-D_0}
e^{i\phi_0} \alpha \beta^*
\\ \mbox{c.c.} &
(1-F_1) |\beta |^2 \end{array} \right) ,
    \label{transf0-rho}\end{equation}
and for the measurement result 1 the transformation is
\begin{eqnarray}
\alpha |0\rangle +\beta |1\rangle \rightarrow &&
 \nonumber \\
 && \hspace{-2.1cm}
 \frac{1}{P_1} \left(
\begin{array}{cc} (1-F_0) |\alpha |^2 \,\, & \sqrt{(1-F_0)F_1}\, e^{-D_1}
e^{i\phi_1} \alpha \beta^*
\\ \mbox{c.c.} &
F_1 |\beta |^2 \end{array} \right) , \quad \,\,
    \label{transf1-rho}\end{eqnarray}
where c.c.\ in the density matrix means complex conjugation of the
opposite off-diagonal element, and the probabilities $P_0$ and $P_1$
of the measurement results are still given by Eq.\ (\ref{P_0,1}).

    Using the quantum mechanics linearity, these equations can be easily
generalized to an arbitrary (mixed) initial state $\rho$. For the
result 0 the qubit density matrix transformation is
\begin{equation}
\left( \begin{array}{cc} \rho_{00}& \rho_{01} \\ \rho_{10} &
\rho_{11}
\end{array} \right) \rightarrow \frac{1}{P_0} \left(
\begin{array}{cc} F_0 \rho_{00}& \sqrt{F_0(1-F_1)}\, e^{-D_0}
e^{i\phi_0} \rho_{01}
\\ \mbox{c.c.} &
(1-F_1) \rho_{11} \end{array} \right) ,
    \label{transf0-rho-2}\end{equation}
and for the result 1 the transformation is
\begin{eqnarray}
\left( \begin{array}{cc} \rho_{00}& \rho_{01} \\ \rho_{10} &
\rho_{11}
\end{array} \right) \rightarrow &&
\nonumber \\
 && \hspace{-2.2cm}
 \frac{1}{P_1} \left(
\begin{array}{cc} (1-F_0) \rho_{00}\,\, & \sqrt{(1-F_0)F_1}\, e^{-D_1}
e^{i\phi_1} \rho_{01}
\\ \mbox{c.c.} &
F_1 \rho_{11} \end{array} \right) , \quad \,\,
    \label{transf1-rho-2}\end{eqnarray}
while the probabilities of the results 0 and 1 are
\begin{equation}
P_{0}=F_0 \rho_{00} +(1-F_1) \rho_{11}, \,\,\, P_{1}=(1-F_0)
\rho_{00} +F_1 \rho_{11}.
    \label{P_0,1-rho}\end{equation}

Equations (\ref{transf0-rho-2})--(\ref{P_0,1-rho}) give the {\it
complete description} of the qubit measurement by a binary-outcome
QND detector. Notice that the fidelities $F_i$  and decoherences
$D_i$ satisfy obvious inequalities $0\leq F_i\leq 1$ and $D_i\geq
0$, while phases $\phi_i$ are defined modulo $2\pi$.

    If the qubit evolution is averaged over the result, then the
transformation becomes
\begin{eqnarray}
&& \hspace{-0.2cm}
\left( \begin{array}{cc} \rho_{00} \,\,\, & \rho_{01} \\
\rho_{10}\,\,\, & \rho_{11}
\end{array} \right) \rightarrow \left(
\begin{array}{cc} \rho_{00} \,\,\,\, & e^{-D_{\rm av}} e^{i\phi_{\rm av}}
\rho_{01}
\\ \mbox{c.c.} \,\,\, &
\rho_{11} \end{array} \right) ,
    \label{transf-av} \\
&&   e^{-D_{\rm av}} e^{i\phi_{\rm av}} =
 \sqrt{F_0(1-F_1)}\, e^{-D_0} e^{i\phi_0}
\nonumber \\
&& \hspace{2.0cm}  + \sqrt{(1-F_0)F_1}\, e^{-D_1} e^{i\phi_1} ,
\qquad \label{rho01av}\end{eqnarray} where $D_{\rm av}$ describes
decoherence of the ensemble of qubits.
 Since $D_{0,1}\geq 0$, we immediately obtain the following
lower bound for the ensemble decoherence:
\begin{equation}
D_{\rm av} \geq D_{\min} = -\ln [\sqrt{F_0(1-F_1)} +
\sqrt{(1-F_0)F_1}].
    \label{Dmin}\end{equation}
This inequality is a counterpart\cite{Av-Sukh} of the inequality
$\Gamma \geq 1/2\tau_m$ for a linear detector.

    Notice that the decoherence bound (\ref{Dmin}) is purely
informational, and it can also be easily derived in the framework of
the quantum Bayesian formalism. \cite{Kor-99,Kor-01,Kor-rev}
Following the derivation of Ref.\ \onlinecite{Kor-99}, we can obtain
the diagonal elements of the post-measurement density matrix for the
measurement outcomes 0 or 1 via the classical Bayes theorem; the
results coincide with the diagonal matrix elements in Eqs.\
(\ref{transf0-rho-2}) and (\ref{transf1-rho-2}). Then using the
inequality $|\rho_{01}|^2\leq \rho_{00}\rho_{11}$ for each
measurement outcome and averaging over the measurement outcomes, we
immediately obtain the inequality (\ref{Dmin}). The inequality
(\ref{Dmin}) can also be obtained following the derivation of Ref.\
\onlinecite{Av-Sukh}.

    It is easy to show that $D_{\min}\geq 0$ since the fidelities
$F_0$ and $F_1$ are between 0 and 1. The value $D_{\min}=0$ is
realized when $F_0+F_1=1$. In this case the measurement does not
give any information about the qubit state, and therefore the
quantum mechanics allows complete absence of the qubit decoherence.

    The quantum efficiency of a QND binary-outcome detector can be
defined in a number of different ways (some figures of merit for the
QND detectors of qubits have been discussed in Ref.\
\onlinecite{Ralph-Wiseman}).
 Similarly to the definition of efficiency
$\eta =\Gamma_{\min}/\Gamma$ for a linear detector, we can define
the quantum efficiency in our case as
    \begin{equation}
    \eta = D_{\min}/D_{\rm av}.
    \label{eta-av}\end{equation}
The quantum efficiency can also be introduced in the spirit of
definitions $\tilde\eta$ and $\tilde {\tilde \eta}$ for linear
detectors (so that a detector is ideal if $D_0=D_1=0$):
 \begin{equation}
\tilde\eta =\frac{D_{\min}}{-\ln [\sqrt{F_0(1-F_1)}\, e^{-D_0} +
\sqrt{(1-F_0)F_1}\, e^{-D_1}]},
    \label{tilde-eta}\end{equation}
        \begin{equation}
\tilde{\tilde\eta} =\frac{-\ln |\sqrt{F_0(1-F_1)} +
\sqrt{(1-F_0)F_1}\, e^{i(\phi_1-\phi_2)}|}{D_{\rm av}} ,
    \label{tilde-tilde-eta}\end{equation}
where $D_{\rm av}$ and $D_{\min}$ are given by Eqs.\ (\ref{rho01av})
and (\ref{Dmin}).
 Notice that in an experiment the phase difference $\phi_0-\phi_1$ can be
relatively easily zeroed by adding the compensating conditional
phase rotation to the qubit after the measurement. The efficiency
(\ref{eta-av}) of such modified detector corresponds to the
definition $\tilde\eta$ of Eq.\ (\ref{tilde-eta}).

  It is also quite meaningful to define separate quantum efficiencies for
each measurement outcome, since realistic detectors can behave very
differently for different outcomes. For example, the detection of
superconducting phase qubits \cite{Cooper,Claudon,Katz} completely
destroys the qubit in the case of measurement result 1. The
binary-outcome detectors of the charge and flux qubits based on
switching or bifurcation
\cite{Vion,Siddiqi-04,Siddiqi-05,Lupascu-07} are also very
asymmetric in a sense that the detector either switches to a
significantly ``excited'' mode or remains relatively ``quiet''.

    There are several possible ways to introduce outcome-dependent
efficiencies $\eta_0$ and $\eta_1$ (to some extent this is a matter
of taste). In this paper we will mostly use the following
definition:
   \begin{equation}
\eta_0 = \frac{D_{\min}}{D_0+D_{\min}} , \quad
    \eta_1 =\frac{D_{\min}}{D_1+D_{\min}}.
    \label{eta-01}\end{equation}
The advantage of this definition is that $\tilde\eta$ is always in
between $\eta_0$ and $\eta_1$, and therefore coincides with them if
$\eta_0=\eta_1$. However, in some cases a more meaningful definition
is
    \begin{equation}
\tilde\eta_0 = \frac{-\ln \sqrt{F_0(1-F_1)} }{D_0-\ln
\sqrt{F_0(1-F_1)}} , \,\,
    \tilde\eta_1 = \frac{-\ln
\sqrt{(1-F_0)F_1} }{D_0-\ln \sqrt{(1-F_0)F_1}},
    \label{tilde-eta-01}\end{equation}
which naturally stems from the form of the off-diagonal matrix
elements in Eqs. (\ref{transf0-rho-2}) and (\ref{transf1-rho-2})
[the tilde sign here has no relation to the tilde signs in Eqs.\
(\ref{tilde-eta}) and (\ref{tilde-tilde-eta})]; it is easy to see
that $\tilde \eta_i\geq \eta_i$. It is also possible to characterize
the outcome-dependent efficiencies directly by $e^{-D_0}$ and
$e^{-D_1}$.

    \section{Several models of detectors}

In this section we discuss several models of QND binary-outcome
detectors (not necessarily realistic) and analyze their quantum
efficiency.

    \subsection{Indirect projective measurement}

    Let us start with an unrealistic but conceptually simple model
of indirect projective measurement. In this model the measured qubit
interacts with another (ancillary) qubit, which is later measured in
the ``orthodox'' projective way. Assume that the ancillary qubit is
initially in the state $|0_a\rangle$ and the interaction leads to
the following entanglement:
\begin{eqnarray}
  (\alpha |0\rangle +\beta |1\rangle)\, |0_a\rangle & \rightarrow
&  \alpha |0\rangle \, (c_{00} |0_a\rangle +c_{10} |1_a\rangle)
    \nonumber \\
&& \hspace{-0.2cm} +  \beta |1\rangle \, (c_{01} |0_a\rangle +c_{11}
|1_a\rangle),
\end{eqnarray}
where $|c_{00}|^2+|c_{10}|^2=|c_{01}|^2+|c_{11}|^2=1$. If the
ancillary qubit is then measured and found in the state
$|0_a\rangle$, the qubit state becomes $(\alpha c_{00}
|0\rangle+\beta c_{01}|1\rangle)/Norm$, while for the measurement
result 1 the qubit state becomes $(\alpha c_{10} |0\rangle+\beta
c_{11}|1\rangle)/Norm$ (here $Norm$ is the normalization which is
easy to find in each case).

    We see that this model exactly corresponds to the ideal model
considered at the beginning of Sec.\ \ref{III}. For such detector
$F_0=|c_{00}|^2$, $F_1=|c_{11}^2|$, $\phi_0=\arg (c_{00}c_{01}^*)$,
$\phi_1=\arg (c_{10}c_{11}^*)$, and there are no extra decoherences:
$D_0=D_1=0$. Therefore, it is an ideal detector in the sense that
\begin{equation}
\eta_0 =\eta_1=1, \,\,\, \tilde\eta=\tilde{\tilde\eta}=1;
\end{equation}
however, the efficiency $\eta$ is less than 100\% if
$\phi_0\neq\phi_1$:
\begin{equation}
\eta =\frac{-\ln [\sqrt{F_0(1-F_1)} + \sqrt{(1-F_0)F_1}]} {-\ln
|\sqrt{F_0(1-F_1)} + \sqrt{(1-F_0)F_1}\, e^{i(\phi_1-\phi_0})|} .
\end{equation}

    \subsection{Linear detector in binary-outcome regime}

    Now let us consider a binary-outcome detector realized by a
linear detector, which output is compared with a certain threshold
to determine if the output falls into the ``result 0'' or ``result
1'' category. We will characterize the linear detector by two levels
of average current $I_0$ and $I_1$ corresponding to the two qubit
states [without loss of generality we assume that the detector
output is the current $I(t)$] and by the spectral density $S$ of the
output white noise. Then the time needed for signal-to-noise ratio
reaching 1 is $\tau_m=2S/(\Delta I)^2$ where $\Delta I=I_1-I_0$.
Since the qubit evolves only due to measurement, the qubit evolution
is described by the quantum Bayesian equations\cite{Kor-99,Kor-01}
    \begin{eqnarray}
 \frac{\rho_{00}(t)}{\rho_{11}(t)} =\frac{\rho_{00}(0)\exp
[-(\bar{I}-I_0)^2t/S]}{\rho_{11}(0)\exp [-(\bar{I}-I_1)^2t/S]} ,
\qquad \qquad
\label{Bayes-1}\\
 \rho_{01}(t)=\rho_{01}(0)
\sqrt{\frac{\rho_{00}(t)\rho_{11}(t)}{\rho_{00}(0)\rho_{11}(0)}} \,
e^{iK(\bar{I}-\frac{I_0+I_1}{2})t} e^{-\gamma t} ,
    \label{Bayes-2}\end{eqnarray}
where the average current $\bar{I}(t)=t^{-1}\int_0^t I(t')\, dt'$
carries all information about the measurement result, $K$ is the
correlation between the output and back-action noise, and
decoherence rate $\gamma$ is related to the ensemble decoherence
rate $\Gamma$ as $\gamma =\Gamma -(\Delta I)^2/4S -K^2 S/4$ [for
simplicity we neglect possible extra factor $e^{i\varepsilon t}$ in
Eq.\ (\ref{Bayes-2}) due to constant energy shift; the effect of
this factor is trivial]. The probability distribution of the result
$\bar{I}$ is
\begin{equation}
    P(\bar{I})=\sum_{i=1,2} \rho_{ii}(0)\sqrt{t/\pi S}\exp [-(\bar{I}-I_i)^2
    t/S].
\end{equation}
   Introducing dimensionless measurement result as $r=[\bar{I}-\frac{I_0+I_1}{2}]\,
\sqrt{t/S}$, and assuming $I_1>I_0$, Eq.\ (\ref{Bayes-1}) can be
rewritten as
    \begin{equation}
\frac{\rho_{00}(t)}{\rho_{11}(t)} =\frac{\rho_{00}(0)}{\rho_{11}(0)}
\, e^{-r\sqrt{8t/\tau_m}},
    \label{Bayes-1-m}\end{equation}
 while the
probability distribution becomes
    \begin{equation}
P(r)=\frac{1}{\sqrt{\pi}}\sum_{i=1,2} \rho_{ii}(0)
e^{-[r+(-1)^i\sqrt{t/2\tau_m}\,]^2}.
    \label{P(r)}    \end{equation}

    Fixing the time of measurement $t$, the binary-outcome detector
can be realized by comparing the result $r$ with a certain threshold
$r_{th}$, so that if $r\geq r_{th}$ the outcome is considered to be
1, otherwise it is considered to be 0. The fidelities of such
detector can be easily calculated:
\begin{equation}
 F_0=\frac{1+\mbox{erf}
(r_{th}+s)}{2}, \,\,\, F_1=\frac{1+\mbox{erf} (-r_{th}+s)}{2}.
    \label{F01-linear}\end{equation}
where $s=\sqrt{t/2\tau_m}$ and
$\mbox{erf}(x)=(2/\sqrt{\pi})\int_0^x e^{-z^2}dz$ is the error
function.

    In the case of the result 0, the resulting density matrix given by Eqs.\
(\ref{Bayes-1-m}) and (\ref{Bayes-2}) should be averaged over $r$
within the range $(-\infty, r_{th})$ with the weight given by Eq.\
(\ref{P(r)}).
    It is easy to check that the obtained diagonal matrix elements
$\rho^{(0)}_{ii}$ coincide with the diagonal elements in Eq.\
(\ref{transf0-rho-2}); this is a trivial fact since the diagonal
matrix elements should obey the classical Bayes formula. For
averaging of the off-diagonal matrix element [Eq.\ (\ref{Bayes-2})]
let us assume for simplicity $K=0$, then we obtain
    \begin{equation}
    \rho_{01}^{(0)} =\frac{e^{-\gamma
    t}e^{-s^2}[1+\mbox{erf}(r_{th})]/2}{\rho_{00}F_0+\rho_{11}(1-F_1)}
    \, \rho_{01}
    \label{rho01-0-linear}\end{equation}
(in this notation $\rho$ denotes pre-measured state, while
$\rho^{(0)}$ denotes post-measurement state corresponding to the
result 0).  Then
    \begin{equation}
    D_0= \gamma t + s^2 -\ln \frac{1+\mbox{erf}(r_{th})}{2\sqrt{F_0(1-F_1)}},
    \label{D0-linear}\end{equation}
and using the definition (\ref{eta-01}) of quantum efficiency
$\eta_0$, we find it as
   \begin{equation}
\eta_0=\left[ 1+\frac{D_0}{-\ln
[\sqrt{F_0(1-F_1)}+\sqrt{(1-F_0)F_1]}}\right]^{-1} .
   \end{equation}

\begin{figure}[tb]
  \centering
 \includegraphics[width=8cm]{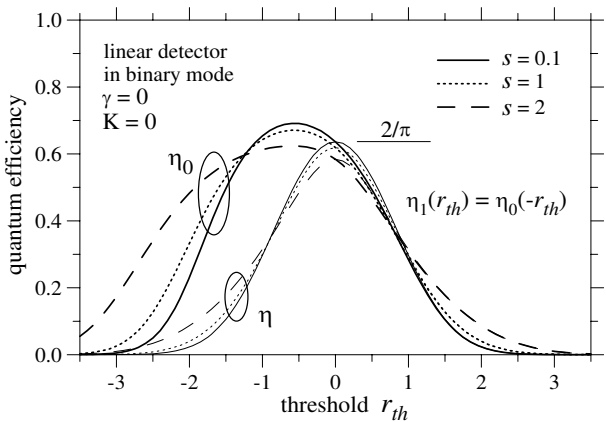}
  \caption{The result-0 quantum efficiency $\eta_0$ (thick lines)
and averaged efficiency $\eta$ (thin lines) for a linear detector in
a binary-outcome mode, as functions of the threshold $r_{th}$
separating results 0 and 1. Solid, dotted, and dashed lines are for
different strengths of measurement: $s=0.1$, 1, and 2,
correspondingly. We assume an ideal linear detector: $\gamma =0$ and
$K=0$. }
  \label{fig1}
\end{figure}

Notice that even for an ideal linear detector ($\gamma =0$) the
quantum efficiency $\eta_0$ is not 100\%. Thick lines in Fig.\
\ref{fig1} show the dependence of $\eta_0$ in this case on the
chosen threshold $r_{th}$ for several values of the parameter
$s=\sqrt{t/2\tau_m}$, which characterizes the measurement strength.
One can see that the curves are not symmetric, and the asymmetry
grows with increase of $s$. The line corresponding to $s=0.1$ (thick
solid line) practically coincides with the result in the limit
$s\rightarrow 0$ (it is easy to derive a formula for this limit;
however, it is long and we do not show it here). As follows from the
numerical results, $\eta_0 < 0.692$ always, and the maximum is
achieved at $s\approx 0$ and $r_{th}\approx -0.563$.

    Analysis of the resulting density matrix in the case of
measurement result 1 is similar to the above analysis. As obvious
from the symmetry, the matrix element $\rho_{01}^{(1)}$ is given by
Eq.\ (\ref{rho01-0-linear}) with $\mbox{erf}(r_{th})$ replaced by
$\mbox{erf}(-r_{th})$ and exchanged fidelities $F_0\leftrightarrow
F_1$ [notice that the transformation $r_{th}\rightarrow -r_{th}$
exchanges the fidelities in Eq.\ (\ref{F01-linear})].
Correspondingly, $D_1$ is given by Eq.\ (\ref{D0-linear}) modified
in the same way, and there is a simple symmetry
    \begin{equation}
\eta_1(r_{th})=\eta_0(-r_{th})
    \end{equation}
 for the quantum efficiencies (with the same $s$). Therefore, in Fig.\
\ref{fig1} the dependences $\eta_1(r_{th})$ can be obtained by
reflection of the thick lines about the axis $r_{th}=0$.

    Now let us consider the result-independent quantum efficiency.
To calculate the efficiency $\eta$ defined by Eq.\ (\ref{eta-av}) we
notice that $D_{\rm av}=\gamma t +s^2 + s^2 (\tau_m K\Delta I/2)^2$,
which is obviously the same as for the linear detector with linear
output, and therefore
    \begin{equation}
\eta =\frac{-\ln [\sqrt{F_0(1-F_1)}+\sqrt{(1-F_0)F_1}]}{\gamma t
+s^2 \left[ 1+ \left(\tau_m K\Delta I/2\right)^2 \right]} \, .
    \label{eta-linear}\end{equation}
In the case $K\neq 0$ the efficiency $\tilde\eta$ [defined by Eq.\
(\ref{tilde-eta})] is given by Eq.\ (\ref{eta-linear}) without the
term proportional to $K$ in the denominator, while the formula for
the efficiency $\tilde{\tilde \eta}$ [defined by Eq.\
(\ref{tilde-tilde-eta})] is quite long. In the case $K=0$ the three
efficiencies obviously coincide:
$\eta=\tilde\eta=\tilde{\tilde\eta}$.

    Thin lines in Fig.\ \ref{fig1} show the dependence $\eta (r_{th})$
for an ideal linear detector ($\gamma=0$, $K=0$) for several values
of the measurement strength $s$. (If $\gamma=0$, but $K\neq 0$, then
these curves show the efficiency $\tilde\eta$.) We see that the
curves are symmetric, and $\eta$ reaches maximum at $r_{th}=0$. The
efficiency at this point increases with decrease of the measurement
strength $s$; however even for $s\rightarrow 0$ we have an upper
bound $\eta \leq 2/\pi$. Notice that $\eta=\eta_0=\eta_1$ at
$r_{th}=0$ because of the symmetry and chosen definition
(\ref{eta-01}) for outcome-dependent efficiencies.

    The main finding of this subsection is that a linear detector in
a binary-outcome regime is never ideal ($\eta <2/\pi$,
$\eta_i<0.7$), even if the linear detector itself is ideal
($\gamma=0$, $K=0$). This is obviously a consequence of the
information loss, which happens when the actual measurement result
$r$ is reduced to only one of two outcomes: 0 ($r<r_{th}$) or 1
($r>r_{th}$). [Notice that the quantum efficiency of a linear
detector in the standard linear-output regime \cite{Kor-rev} is
given by Eq.\ (\ref{eta-linear}) with the numerator replaced by
$s^2$.]

    \subsection{Detector of the superconducting phase qubit}

\begin{figure}[tb]
  \centering
 \includegraphics[width=7cm]{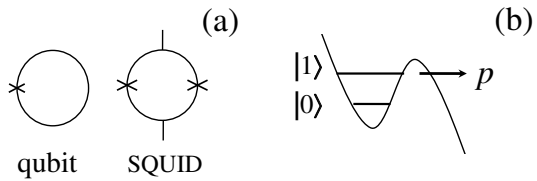}
  \caption{(a) Schematic of a superconducting phase qubit coupled
to a SQUID. (b) Energy profile of the qubit with two lowest energy
levels in the well representing logic states $|0\rangle$ and
$|1\rangle$. Measurement is performed by lowering the energy
barrier, so that the state $|1\rangle$ can tunnel out of the well
with probability $p$; the tunneling event is then sensed by the
SQUID.}
  \label{fig2}
\end{figure}

    So far there is only one direct experiment showing high
quantum efficiency of a binary-outcome detector of a solid-state
qubit. This is the experiment on partial collapse of the
superconducting phase qubit. \cite{Katz} In this experiment the
qubit is made of a superconducting loop interrupted by a Josephson
junction [see Fig.\ \ref{fig2}(a)]; the corresponding potential
profile is shown in Fig.\ \ref{fig2}(b). Two lowest energy levels in
the quantum well represent the logic states $|0\rangle$ and
$|1\rangle$. The qubit is measured \cite{Cooper} by reducing the
barrier of the quantum well (by changing the magnetic flux through
the loop), so that the state $|1\rangle$ can tunnel out of the well,
while the state  $|0\rangle$ does not tunnel out. The tunneling
event or its absence is checked at a later time by using an extra
SQUID [Fig.\ \ref{fig2}(a)], which is off when the qubit barrier is
lowered, and therefore does not affect the tunneling process.

    By varying the amplitude and duration of the measurement pulse
which lowers the barrier, it is possible to control the probability
$p$ of tunneling from the level $|1\rangle$, which characterizes the
measurement strength. [For a rectangular pulse $p=1-e^{-\Gamma t}$,
where $\Gamma$ is the tunneling rate and $t$ is the pulse duration.]
Neglecting all imperfections (including finite tunneling from the
state $|0\rangle$), the fidelities of such measurement are
    \begin{equation}
F_0=1, \qquad  F_1= p .
    \label{fidel-phase}\end{equation}
In the case of measurement result 1 (registered tunneling event) the
qubit state is completely destroyed (no longer in the quantum well).
However, for measurement result 0 (null-result, no tunneling) the
system remains in the quantum well, and therefore it is meaningful
to discuss the qubit state evolution due to measurement. Ideally,
this evolution should be given by Eq.\ (\ref{transf0}), and this is
exactly what has been confirmed in the experiment \cite{Katz} with
good accuracy.

    Since the qubit state is destroyed for the measurement result 1,
the quantum efficiency $\eta$ cannot be defined, as well as the
efficiencies $\tilde\eta$, $\tilde{\tilde\eta}$, and $\eta_1$.
However, the null-result efficiency $\eta_0$ is a well-defined
quantity. In the ideal case described by Eq.\ (\ref{transf0}) the
measurement does not dephase the qubit state, and therefore
    \begin{equation}
    \eta_0=1.
    \end{equation}
In a realistic case there are always some mechanisms, which lead to
the qubit decoherence via processes of virtual tunneling, and
correspondingly decrease $\eta_0$. Some of these processes have been
considered theoretically in Ref.\ \onlinecite{Pryadko}, and the
results of that paper for the null-result qubit decoherence can be
converted into the results for the efficiency $\eta_0$.

    To estimate experimental quantum efficiency $\eta_0$, we use Fig.\
3(c) of Ref.\ \onlinecite{Katz} (notice that $\tilde\eta_0=\eta_0$
because $F_0=1$). Choosing the data for the initial state
$(|0\rangle+|1\rangle )/\sqrt{2}$ and moderate measurement strength
($p \simeq 0.5$), we see that the process of measurement reduces the
visibility of the tomography oscillations by less than 7\% (we
exclude the effects of energy relaxation and dephasing, which occur
even in the absence of measurement). Since the theoretical
visibility is
$[1-4\rho_{00}^{(0)}\rho_{11}^{(0)}(1-e^{-2D_0})]^{1/2}$, and since
$\rho_{00}^{(0)}\rho_{11}^{(0)}=2/9$ for our initial state and
$p=0.5$, we estimate dephasing as $D_0\alt 0.08$. Using
$D_{\min}=-(1/2)\ln (1-p)=\ln 2/2$, we finally convert $D_0$ into
the quantum efficiency $\eta_0 \agt 0.8$. Notice that this result
most likely underestimates $\eta_0$; if we use the decoherence value
of 4\% obtained in Ref.\ \onlinecite{Katz} in a different way, then
the quantum efficiency $\eta_0$ is over 90\%.

    Notice that in the experiment of Ref.\ \onlinecite{Katz} the
detection of the tunneling event by SQUID was done after the
tomography pulse sequence, and therefore did not affect the quantum
efficiency of the partial measurement. If the detection by SQUID
should be included into the partial measurement protocol, then it is
important to avoid the qubit decoherence by the SQUID operation.
This decoherence can be significantly decreased by using the SQUID
in the null-result mode also. If the SQUID exceeds its critical
current earlier for the tunneled qubit than for the non-tunneled
qubit, and if the SQUID is biased in between these two critical
currents, then detection of the tunneling event is still accurate;
however, in absence of the qubit tunneling the SQUID remains in the
``quiet'' S-state.

    Now let us briefly discuss the natural generalization\cite{Pryadko} of
the null-result measurement of the phase qubit to the case when
there is a non-zero probability $p_0$ of the qubit tunneling from
the state $|0\rangle$. In this case the measurement fidelities are
    \begin{equation}
F_0= 1-p_0, \qquad F_1 =p.
    \end{equation}
 Assuming that the coupling between the
two qubit states due to tunneling is negligible, we can still use
Eq.\ (\ref{transf0}) to describe the null-result
evolution.\cite{Pryadko} In this case the detector is still ideal in
the sense that $\eta_0=1$ (while $\eta_1$, $\eta$, $\tilde{\eta}$,
and $\tilde{\tilde\eta}$ are still not defined). Notice that for
non-zero dephasing $D_0$ the definition (\ref{eta-01}) for $\eta_0$
differs from the definition (\ref{tilde-eta-01}) for $\tilde\eta_0$,
in contrast to the above case of $p_0=0$, when the two definitions
coincide.

    \subsection{Tunneling-into-continuum detector}

\begin{figure}[tb]
  \centering
\includegraphics[width=5.5cm]{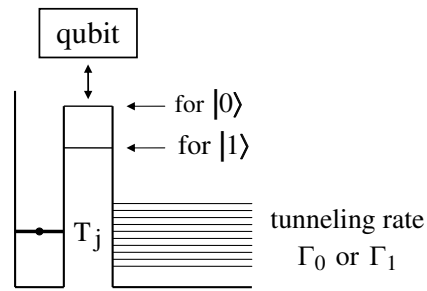}
  \caption{Schematic of a detector based on tunneling into continuum.
The qubit state controls the tunneling matrix element ($T_0$ or
$T_1$) and therefore the tunneling rate ($\Gamma_0$ or $\Gamma_1$).
}
  \label{fig3}
\end{figure}

    An obvious drawback of the detector considered in the previous
subsection is the fact that the qubit state is completely destroyed
when the measurement result is 1. In this subsection we consider a
detector, which is still based on tunneling into continuum; however,
it does not destroy the qubit state for both measurement results.
The schematic of the detector is shown in Fig.\ \ref{fig3}. The
initial state of the detector is in the quantum well, and it can
tunnel through a barrier into continuum  (the phase space is
arbitrary). The barrier height is modulated by the qubit state, so
that the states $|0\rangle$ and $|1\rangle$ correspond to different
rates of tunneling: $\Gamma_0$ and $\Gamma_1$ (we assume $\Gamma_1 >
\Gamma_0$). The measurement is performed during a finite time $t$,
after which it is checked if the tunneling has occurred (result 1)
or not (result 0). The measurement fidelities are obviously
    \begin{equation}
    F_0=\exp (-\Gamma_0 t), \qquad F_1= 1-\exp(-\Gamma_1 t),
    \label{tun-fid}\end{equation}
and the goal of this subsection is to analyze the quantum
efficiencies of such detector.

    The main difference of this detector compared to the
detector discussed in the previous subsection is that the tunneling
happens in a physical system different from the qubit, and therefore
the qubit state is not destroyed by the measurement. As a price for
this improvement, the detector now requires two stages: a ``sensor''
which can tunnel, and then detection of the tunneling event, while
for the previous detector the tunneling sensor was physically
combined with the qubit. Notice that the model we consider has some
similarity with the bifurcation detectors,
\cite{Siddiqi-05,Lupascu-07,Dykman-80} which are used for the
measurement of qubits (though there are significant differences as
well\cite{Dykman-07}).

We describe the qubit-detector system by the following Hamiltonian:
    \begin{eqnarray}
&&    H=\sum_k \varepsilon_k |k\rangle \langle k|+ |0\rangle \langle
0| \sum_k (T_{0,k} |k\rangle \langle w| + T_{0,k}^* |w\rangle
\langle k|)
\nonumber \\
&& \hspace{0.7cm}     + |1\rangle \langle 1| \sum_k (T_{1,k}
|k\rangle \langle w|    + T_{1,k}^* |w\rangle \langle k|),
    \label{Hamilt}\end{eqnarray}
where the detector Hilbert space consists of the state $|w\rangle$
in the well (its energy is taken to be zero) and many energy levels
$|k\rangle$ (with energies $\varepsilon_k$) representing the
continuum. Since we assume the QND measurement, the qubit
Hamiltonian is zero (if energies of states $|0\rangle$ and
$|1\rangle$ are actually different, the qubit Hamiltonian is still
zero in the rotating frame). The coupling with the qubit changes the
detector tunneling matrix elements from $T_{0,k}$ for the qubit
state $|0\rangle$ to  $T_{1,k}$ for the qubit state $|1\rangle$.

    Assuming that the qubit is in one of the logic  states $|j\rangle$ ($j=0,1$),
the evolution of the detector wavefunction $a_j(t) |w\rangle +\sum_k
b_{j,k}(t) |k\rangle$ is given by equations
    \begin{equation}
    \dot{a}_j=-\frac{i}{\hbar}\sum_k T_{j,k}^* b_{j,k}, \quad
    \dot{b}_{j,k}= -\frac{i}{\hbar} \varepsilon_k -\frac{i}{\hbar}
    T_{j,k} a_j
    \end{equation}
with the initial condition $a_j(0)=1$, $b_{j,k}(0)=0$. Assuming the
simplest case when $T_{j,k}=\mbox{const} =T_j$ and the energy levels
are very dense with constant density of states ${\cal D}$ and
infinite energy bandwidth, we obtain the standard solution (see
Ref.\ \onlinecite{Pryadko} for better approximations)
    \begin{equation}
    a_j(t) = e^{-\Gamma_j t/2}, \quad
      b_{j,k}(t) = \frac{-i}{\hbar} \, T_j \, \frac{e^{-\Gamma_j t/2}
    -e^{-i\varepsilon_k t/\hbar}}{-\Gamma_j/2
    +i\varepsilon_k/\hbar},
     \label{a-b-solution}\end{equation}
where $ \Gamma_j = (2\pi /\hbar) |T_j|^2{\cal D}$.

    If the initial state of the qubit is $\alpha |0\rangle +\beta
|1\rangle$ (where $|\alpha|^2+|\beta|^2=1$), the evolution is given
by superposition of the two evolutions:
\begin{equation}
(\alpha |0\rangle a_0 + \beta |1\rangle a_1 ) |w\rangle
+{\small\sum}_k (\alpha |0\rangle b_{0,k} + \beta |1\rangle b_{1,k})
|k\rangle .
    \label{tun-superp}\end{equation}
 When after time $t$ it is checked if the tunneling
event has occured or not, the corresponding probabilities of the
measurement outcomes can be obtained by squaring the coefficients in
Eq.\ (\ref{tun-superp}):
    \begin{equation}
P_0=|\alpha|^2 e^{-\Gamma_0 t}+ |\beta|^2 e^{-\Gamma_1 t}, \quad
P_1=1-P_0.
    \end{equation}
These formulas are obviously consistent with the general description
(\ref{P_0,1}) and fidelities (\ref{tun-fid}).

    In the case of the measurement result 0 (no tunneling), the
state (\ref{tun-superp}) gets projected onto the subspace, which
contains only the vector $|w\rangle$ from the detector degrees of
freedom; therefore the overall evolution due to measurement is
         \begin{equation}
(\alpha |0\rangle +\beta |1\rangle ) \, |w\rangle \rightarrow
 \frac{\alpha |0\rangle \,
e^{-\Gamma_0 t/2}+\beta |1\rangle \, e^{-\Gamma_1
t/2}}{\sqrt{|\alpha|^2 e^{-\Gamma_0 t}+ |\beta|^2 e^{-\Gamma_1 t}}}
\, |w\rangle,
    \label{tun-res0}\end{equation}
where the denominator is due to normalization and is obviously equal
to $\sqrt{P_0}$, as in the general description (\ref{transf0}).
Notice that in Eq.\ (\ref{tun-res0}) the qubit state is not
entangled with the detector [even though it was entangled in the
process -- see Eq.\ (\ref{tun-superp})], and therefore we can say
that the qubit state remained pure and underwent a coherent
non-unitary evolution described by Eq.\ (\ref{tun-res0}) without
$|w\rangle$. Then considering an arbitrary initial state $\rho$ of
the qubit as a mixture of pure states, it is easy to find that the
after-measurement state is
     \begin{equation}
\rho^{(0)} =      \frac{1}{P_0} \left(
\begin{array}{cc} e^{-\Gamma_0 t} \rho_{00}\quad
& e^{-(\Gamma_0+\Gamma_1) t/2} \rho_{01}
\\ e^{-(\Gamma_0+\Gamma_1) t/2} \rho_{10} \quad &
e^{-\Gamma_1 t} \rho_{11} \end{array} \right) ,
    \end{equation}
where $P_0=\rho_{00}e^{-\Gamma_0 t}+\rho_{11}e^{-\Gamma_1 t}$.
 The corresponding quantum efficiency is obviously ideal:
    \begin{equation}
    \eta_0 = 1,
    \end{equation}
since $D_0=0$.

    In the case of the measurement result 1 (tunneling event) the
wavefunction (\ref{tun-superp}) is projected onto the subspace
orthogonal to $|w\rangle$ and becomes
    \begin{eqnarray}
&& \frac{1}{\sqrt{P_1}} \sum_k \left( \alpha |0\rangle
\frac{-i}{\hbar} \, T_0 \, \frac{e^{-\Gamma_0 t/2}
    -e^{-i\varepsilon_k t/\hbar}}{-\Gamma_0/2
    +i\varepsilon_k/\hbar} \right.
  \nonumber \\
&& \hspace{0.5cm} \left.     + \beta |1\rangle \frac{-i}{\hbar} \,
T_1 \, \frac{e^{-\Gamma_1 t/2}
    -e^{-i\varepsilon_k t/\hbar}}{-\Gamma_1/2
    +i\varepsilon_k/\hbar} \right) |k\rangle , \qquad
    \label{tun-wf-1}\end{eqnarray}
where the factor $1/\sqrt{P_1}$ is again due to normalization. If we
want to discuss only the qubit evolution, we have to convert this
state into a density matrix and then trace over the detector states
$|k\rangle$. It is easy to find that the diagonal matrix elements of
thus obtained qubit density matrix $\rho^{(1)}$ are
$\rho_{00}^{(1)}=|\alpha|^2\sum_k|b_{0,k}|^2/P_1=|\alpha|^2(1-e^{-\Gamma_0
t})/P_1$ and
$\rho_{11}^{(1)}=|\beta|^2\sum_k|b_{1,k}|^2/P_1=|\beta|^2(1-e^{-\Gamma_1
t})/P_1$. They are obviously the same as in the general equation
(\ref{transf1-rho}), since they should obey the classical Bayes
formula. To find the off-diagonal matrix element
$\rho_{01}^{(1)}=\alpha\beta^*\sum_k b_{0,k} b_{1,k}^*/P_1$,  we
perform integration over the energy $\varepsilon$ using the residue
theorem and obtain
    \begin{equation}
    \rho_{01}^{(1)} =\frac{\alpha \beta^*}{P_1} \,
    \frac{ 4\pi T_0 T_1^*\, {\cal D}}{\hbar \, (\Gamma_0+\Gamma_1)}
    \, [1-e^{-(\Gamma_0+\Gamma_1)t/2}] .
    \end{equation}
Finally, expressing $|T_0|$ and $|T_1|$ via $\Gamma_0$ and
$\Gamma_1$, and considering arbitrary initial qubit state $\rho$, we
find the post-measurement qubit state as
    \begin{equation}
    \rho^{(1)}= \frac{1}{P_1} \left(
\begin{array}{cc} (1-e^{-\Gamma_0 t}) \rho_{00}&
\frac{2\sqrt{\Gamma_0\Gamma_1}}{\Gamma_0+\Gamma_1} e^{i\phi_1}
(1-e^{-\frac{\Gamma_0+\Gamma_1}{2} t}) \rho_{01}
\\ \mbox{c.c.} &
(1- e^{-\Gamma_1 t}) \rho_{11} \end{array} \right)
    \end{equation}
where $P_1=\rho_{00}(1-e^{-\Gamma_0 t})+\rho_{11}(1-e^{-\Gamma_1
t})$ and $\phi_1=\arg (T_0/T_1)$. Comparing this result with Eq.\
(\ref{transf1-rho-2}), we find non-zero decoherence
    \begin{equation}
D_1 = -\ln \left[ \frac{2\sqrt{\Gamma_0\Gamma_1}}{\Gamma_0+\Gamma_1}
\,\frac{1-e^{-(\Gamma_0+\Gamma_1)t/2}}{\sqrt{(1-e^{-\Gamma_0
t})(1-e^{-\Gamma_1 t})}}
 \right].
    \label{tun-d1}\end{equation}
Since the averaged informational decoherence bound (\ref{Dmin}) is
     \begin{equation}
     D_{\min}= -\ln \left[ \sqrt{e^{-\Gamma_0 t}e^{-\Gamma_1 t}} +
     \sqrt{(1-e^{-\Gamma_0 t})(1-e^{-\Gamma_1 t})} \right] ,
     \label{tun-dmin}\end{equation}
the quantum efficiency $\eta_1$ defined by Eq.\ (\ref{eta-01}) is

   \begin{equation}
\eta_1 =\left[1+\frac{\ln \left[
\frac{2\sqrt{\Gamma_0\Gamma_1}}{\Gamma_0+\Gamma_1}
\,\frac{1-e^{-(\Gamma_0+\Gamma_1)t/2}}{\sqrt{(1-e^{-\Gamma_0
t})(1-e^{-\Gamma_1 t})}}
 \right]}{\ln [ \sqrt{e^{-(\Gamma_0+\Gamma_1) t}} +
     \sqrt{(1-e^{-\Gamma_0 t})(1-e^{-\Gamma_1 t})} ]}
     \right]^{-1} .
     \end{equation}

\begin{figure}[tb]
  \centering
\includegraphics[width=8cm]{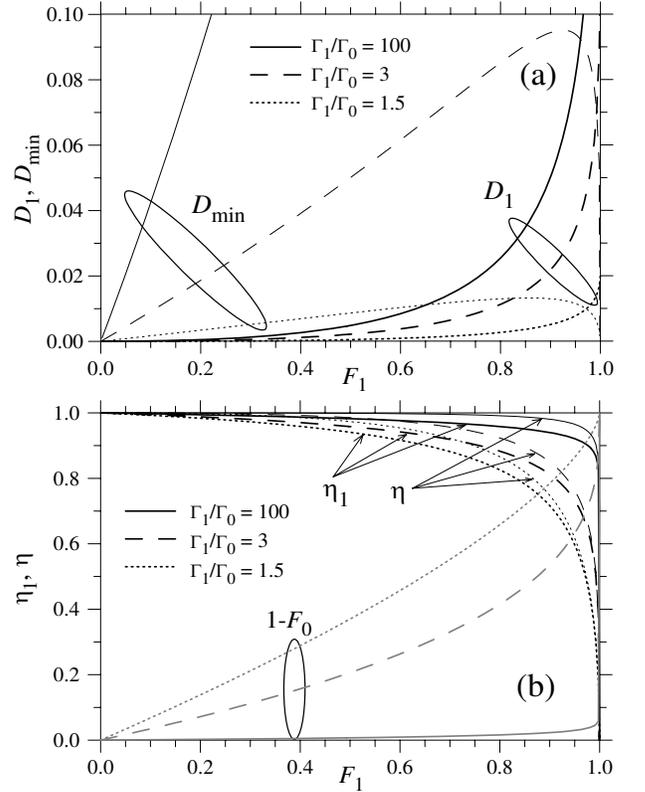}
  \caption{(a) The decoherences $D_1$ (thick lines) and $D_{\min}$ (thin lines)
and (b) the quantum efficiencies $\eta_1$ (thick lines) and $\eta$
(thin lines), as functions of the fidelity $F_1$ for a
tunneling-into-continuum detector with several values of the ratio
of tunneling rates: $\Gamma_1/\Gamma_0=100$ (solid lines), 3 (dashed
lines), and 1.5 (dotted lines). The values of $1-F_0$ are shown by
gray lines in (b). For the calculation of $\eta$ we have assumed
$\phi_1=0$. For null-result outcome the detector is ideal: $D_0=0$,
$\eta_0=1$. }
  \label{fig4}
\end{figure}

    Thick lines in Fig.\ \ref{fig4}(a) show the dependence of the decoherence
$D_1$ on the fidelity $F_1$ for several ratios of the tunneling
rates $\Gamma_1/\Gamma_0$. [The corresponding values of $1-F_0$ are
shown by gray lines in Fig.\ \ref{fig4}(b).] The quantum
efficiencies $\eta_1$ are shown by thick lines in Fig.\
\ref{fig4}(b) for the same parameters, and $D_{\min}$ are shown by
thin lines in Fig.\ \ref{fig4}(a). One can see that $D_1$ approaches
zero and $\eta_1$ approaches 100\% when $F_1\rightarrow 0$
(correspondingly $F_0\rightarrow 1$). This behavior can be
understood from the result for $b_{j,k}(t)$ given by Eq.\
(\ref{a-b-solution}). It is easy to check that in the case $\Gamma_1
t\ll 1$ (then $\Gamma_0 t\ll 1$ also) this result reduces to
$b_{j,k}=(-T_j/\varepsilon_k)(1- e^{-i\varepsilon_k t/\hbar})$.
Since in this case the shape of $b_{j,k}$ dependence on $\epsilon_k$
does not depend on $T_j$, the qubit state in Eq.\ (\ref{tun-wf-1})
becomes disentangled from the detector state, and as a result, the
qubit state remains pure. In contrast, when $\Gamma_1 t  \agt 1$,
the difference between the shapes of $b_{0,k}$ and $b_{1,k}$
contains an information about the qubit state, which is lost since
we do not measure $b_k$; as a result, there is a non-zero
decoherence of  the qubit state. Increase of this lost information
with $\Gamma_1 t$ explains increase of $D_1$ and decrease of
$\eta_1$ with $F_1$ in Figs.\ \ref{fig4}(a) and \ref{fig4}(b). In
the limiting case $F_1\rightarrow 1$ when $\Gamma_1 t > \Gamma_0 t
\gg 1$, the decoherence saturates: $D_1 \rightarrow -\ln
[2\sqrt{\Gamma_0\Gamma_1}/(\Gamma_0+\Gamma_1)]$, while $D_{\min}$
(which describes the information) continues to decrease:
$D_{\min}\approx e^{-\Gamma_0 t}/2$. Since our definition of
$\eta_1$ is based on comparing $D_1$ with $D_{\min}$, this leads to
$\eta_1\rightarrow 0$, as seen in Fig.\ \ref{fig4}(b). It is
interesting to notice that while the decoherence $D_1$ increases
with the increase of the ratio $\Gamma_1/\Gamma_0$ for a fixed
$F_1$, the efficiency $\eta_1$ also increases (instead of
decreasing); this is because $D_{\min}$ increases with
$\Gamma_1/\Gamma_0$ faster than $D_1$.

For the alternative definition (\ref{tilde-eta-01}) the
outcome-dependent quantum efficiencies are
\begin{equation}
\tilde\eta_0=1, \quad
     \tilde\eta_1 = \frac{\ln \sqrt{(1-e^{-\Gamma_0
t})(1-e^{-\Gamma_1 t})} }{\ln \left[
\frac{2\sqrt{\Gamma_0\Gamma_1}}{\Gamma_0+\Gamma_1}
\,[1-e^{-(\Gamma_0+\Gamma_1)t/2}]
 \right]}  .
\end{equation}

   If the resulting qubit state is averaged over the measurement
results 0 and 1, then the averaged density matrix $\rho^{({\rm
av})}=P_0\rho^{(0)}+P_1\rho^{(1)}$ is
    \begin{eqnarray}
    \rho^{({\rm av})}= \left(
\begin{array}{cc} \rho_{00}\quad &  e^{-D_{\rm av}}e^{i\phi_{\rm av}}  \rho_{01}
\\ \mbox{c.c.} \quad &
 \rho_{11} \end{array} \right) , \qquad\qquad  &&
    \\
 e^{-D_{\rm av}}e^{i\phi_{\rm av}} = e^{-\frac{\Gamma_0+\Gamma_1}{2} t} +
\frac{2\sqrt{\Gamma_0\Gamma_1}}{\Gamma_0+\Gamma_1} e^{i\phi_1}
(1-e^{-\frac{\Gamma_0+\Gamma_1}{2} t}). \,\,\,\, &&
    \end{eqnarray}
The quantum efficiencies $\eta$, $\tilde{\eta}$, and
$\tilde{\tilde\eta}$ can then be calculated using Eqs.\
(\ref{eta-av})--(\ref{tilde-tilde-eta}) (notice that in our model
$\phi_0=0$). In particular, when $\phi_1=0$ (the qubit does not
change the phase of tunneling coefficients) these three efficiencies
coincide and are equal to
\begin{equation}
\eta = \frac{\ln \left[ e^{-\frac{\Gamma_0+\Gamma_1}{2} t} +
     \sqrt{(1-e^{-\Gamma_0 t})(1-e^{-\Gamma_1 t})} \right]}
     {\ln \left[ e^{-\frac{\Gamma_0+\Gamma_1}{2} t} +
\frac{2\sqrt{\Gamma_0\Gamma_1}}{\Gamma_0+\Gamma_1}
(1-e^{-\frac{\Gamma_0+\Gamma_1}{2} t}) \right]}
    \label{tun-eta}\end{equation}
(the efficiency $\tilde\eta$ is given by this expression even if
$\phi_1\neq 0$).
    Thin lines in Fig.\ \ref{fig4}(b) show the quantum efficiency $\eta$
given by Eq.\ (\ref{tun-eta}) for the same parameters as for
$\eta_1$. One can see that $\eta >\eta_1$; this is because
$\eta_0=1$ and for our definitions (\ref{tilde-eta}) and
(\ref{eta-01}) the value of $\tilde{\eta}$ is always in between
$\eta_0$ and $\eta_1$.

   As follows from Eqs.\ (\ref{tun-d1}), (\ref{tun-dmin}), and
(\ref{tun-eta}), in the limiting case when there is no tunneling if
the qubit is in the state $|0\rangle$ ($\Gamma_0\rightarrow 0$,
$\Gamma_1/\Gamma_0\rightarrow \infty$), the results are
\begin{equation}
\eta =1, \, D_{\min}=\frac{\Gamma_1 t}{2}  , \, D_1=-\ln [
\frac{2}{\sqrt{\Gamma_1 t}} \, \frac{1-e^{-\Gamma_1
t/2}}{\sqrt{1-e^{-\Gamma_1 t}}} ],
\end{equation}
so despite $\eta_1\neq 1$, the detector is ideal in the sense
$\eta=1$. This happens because in this case for the measurement
result 1 the qubit is fully collapsed onto the state $|1\rangle$, so
the factor $\sqrt{(1-F_0)F_1}$ in Eq.\ (\ref{transf1-rho-2}) is
zero, and additional dephasing due to $D_1$ does not matter. Notice
that in this case $\tilde\eta_1=1$, that illustrates the usefulness
of the definition (\ref{tilde-eta-01}).

    The main finding of this subsection is that the
tunneling-into-continuum detector is ideal for the result 0 (i.e.\
$\eta_0=1$); but it is in general non-ideal for the measurement
result 1 ($\eta_1<1$), leading to non-ideal averaged efficiency
($\eta <1$). However, the numerical results in Fig.\ \ref{fig4}(b)
show that the quantum efficiencies $\eta_1$ and $\eta$ are typically
rather close to 100\%. [The efficiency $\tilde\eta_1$ (not shown) is
significantly closer to 1 than $\eta_1$ and is even higher than
$\eta$ for not too large ratio $\Gamma_1/\Gamma_0$.]

\section{Conclusion}

    In this paper we have discussed possible ways to introduce the
notion of quantum efficiency for binary-outcome detectors of
solid-state qubits. We consider detectors with imperfect measurement
fidelities (non-projective measurement) and define the quantum
efficiency by comparing the qubit dephasing with the
information-related non-unitary evolution dictated by the quantum
mechanics.

   Our attempt to introduce the quantum efficiency for an arbitrary
binary-outcome detector has failed, because the efficiency should in
general be characterized by 18 parameters, that is obviously
impractical. (The number 18 is the difference between 28 parameters
necessary to describe a general binary-outcome detector and 10
parameters necessary to describe an ideal detector, which does not
decohere the qubit for each measurement result.)

    However, the situation is much simpler for a QND detector. Its
operation can be fully characterized by only 6 parameters (instead
of 28): fidelity $F_i$, phase shift $\phi_i$, and decoherence $D_i$
for each measurement result ($i=0,1$) -- see Eqs.\
(\ref{transf0-rho-2})--(\ref{P_0,1-rho}). Therefore it is not
difficult to introduce a meaningful definition for the quantum
efficiency via a combination of these 6 parameters. However, it can
be done in a variety of ways. By comparing the averaged qubit
decoherence with the informational bound (\ref{Dmin}), we have
introduced three slightly different definitions: $\eta$,
$\tilde\eta$, and $\tilde{\tilde\eta}$ [see Eqs.\
(\ref{eta-av})--(\ref{tilde-tilde-eta})], which are counterparts of
the definitions\cite{Kor-01,Kor-rev} of the quantum efficiency for a
linear detector. We have also introduced outcome-dependent quantum
efficiencies $\eta_i$ [see Eq.\ (\ref{eta-01})] by comparing the
decoherences $D_i$ with the informational bound (\ref{Dmin}).
[Another meaningful way to introduce the outcome-dependent
efficiencies is via Eq.\ (\ref{tilde-eta-01}).] Notice that all
these definitions are not applicable in the ``orthodox'' case of
perfect measurement fidelity: $F_0=F_1=1$.

    After introducing the definitions for the quantum efficiency, we
have calculated the efficiencies for several simple models of a
binary-outcome detector. As follows from the results, it is not easy
to find a model for a practical binary-outcome detector which would
have theoretically perfect quantum efficiency (in contrast to linear
detectors, for which QPC realizes the perfect case). Out of the
models we have considered, the perfect efficiency is realized only
in the indirect projective measurement, when the qubit interacts
with another fully coherent two-level system, which is actually
measured. While the quantum efficiency for such measurement setup is
ideal, it is not a quite practical setup.

   Analyzing a linear detector in the binary-outcome regime (when measurement
result is compared with a threshold), we have found that such
detector cannot have perfect quantum efficiency: $\eta\leq 2/\pi$,
$\eta_{0,1} < 0.7$. The tunneling-based partial measurement of
superconducting phase qubits is theoretically ideal for the
null-result outcome: $\eta_0=1$; however, the qubit state is
destroyed in the case of the measurement result 1, and therefore
quantum efficiencies $\eta_1$ and $\eta$ cannot be defined. We have
also considered a detector based on tunneling into continuum, which
tunneling rate depends on the qubit state. For such a detector the
null-result efficiency is also perfect:  $\eta_0=1$, and even though
the efficiencies $\eta_1$ and $\eta$ are not perfect, their values
can be rather close to 100\%.

Our results hint that the practical binary-outcome detectors of
solid-state qubits available at present (e.g.\ bifurcation
detectors\cite{Vion,Siddiqi-04,Siddiqi-05,Lupascu-07} or the
balanced comparator\cite{Walls-07}) cannot closely approach 100\%
quantum efficiency for both measurement results, even theoretically.
However, this is not a rigorous conclusion, and a more detailed
analysis of the quantum efficiencies of the particular practical
detectors is surely interesting and important.

This work was supported by NSA and IARPA under ARO grant
W911NF-04-1-0204.



\end{document}